\documentclass[pre,twocolumn,showpacs]{revtex4-1}
\usepackage{graphicx}
\usepackage{amsmath,amssymb}
\newcommand{\myvec}[1]{{\mathbf{#1}}}
\newcommand{\rvec}{\myvec{r}}

\newcommand{\hatrvec}{{\hat{\mathbf{r}}}}
\newcommand{\Fvec}{\myvec{F}}

\newcommand{\rhobar}{{\overline\rho}}
\newcommand{\wrho}{w_\rho}

\newcommand{\latin}[1]{{\itshape #1}}

\newcommand{\ie}{\latin{i.\,e.}}
\newcommand{\etal}{\latin{et al.}}

\begin{document}

\title{No-go theorem in many body dissipative particle dynamics}

\author{Patrick B. Warren}

\email{patrick.warren@unilever.com}

\affiliation{Unilever R\&D Port Sunlight, Quarry Road East, Bebington,
  Wirral, CH63 3JW, UK.}

\date{March 11, 2013}

\begin{abstract}
Many body dissipative particle dynamics (MDPD) is a particle-based
simulation method in which the interaction potential is a sum of self
energies depending on locally-sampled density variables.  This
functional form gives rise to density-dependent pairwise forces,
however not all such force laws are derivable from a potential and the
integrability condition for this to be the case provides a strong
constraint.  A strategy to assess the implications of this
constraint is illustrated here by the derivation of a useful no-go
theorem for multicomponent MDPD.
\end{abstract}


\maketitle

Dissipative particle dynamics (DPD) has attracted a lot of interest in
its possibilities for modelling soft condensed matter \cite{FS02,
  NMW03}.  It is characterised by pairwise soft repulsive forces with
a pairwise momentum-conserving thermostat \cite{HK92-EW95}.  Conceived
somewhat later, many-body dissipative particle dynamics (MDPD) holds
much promise as a second generation method \cite{PF01-ER03, TNM02,
  MP07, War0103, MBP08, APL+11, GEM13, SAM+13}, although it has mainly been
applied to vapour-liquid coexistence and free surface simulations
\cite{War0103, MBP08, GEM13, APL+11, SAM+13}.  In the present note I
explore the consequences for MDPD of allowing an arbitrary local
density dependence into the pairwise repulsive forces.  In particular
if one requires that a potential exists, so that the forces are
conservative, the allowed functional form of the pairwise forces is
severely constrained.  I will here outline a strategy for
assessing the implications of this constraint, illustrated by the
generation of a no-go theorem for a multicomponent MDPD force law that
has been suggested in the literature \cite{APL+11, SAM+13}.  The
present result also demonstrates that it is by far better to proceed
from the potential to the forces, rather than the other way around.

Leaving aside the DPD thermostat, which has been well described
elsewhere \cite{GW97}, standard DPD is characterised by pairwise
repulsive forces of the form
\begin{equation}
\Fvec_{ij}=A\,w_C(r_{ij})\,\hatrvec_{ij}
\label{eq:dpd}
\end{equation}
where $\Fvec_{ij}$ is the force acting between the $i$th and $j$th
particles (at positions $\rvec_i$ and $\rvec_j$), $A$ is a repulsion
amplitude, $w_C(r)$ is a weight function, $r_{ij}=|\rvec_j-\rvec_i|$
is the spatial separation, and $\hatrvec_{ij} = (\rvec_j-\rvec_i) /
r_{ij}$ is a unit vector along the line of centres.  I shall assume
the weight function has compact support (\ie\ $w_C=0$ for $r>r_c$
where $r_c$ is a cut-off distance) but for the present purposes it is
not necessary to specify the exact functional form.
 
In contrast, MDPD starts from a rather different viewpoint.  The
potential energy in MDPD is a sum of density dependent one-body terms
\cite{PF01-ER03, War0103, MBP08, TNM02, MP07},
\begin{equation}
U(\{\rvec_i\})={\textstyle\sum_{i}}\, u(\rhobar_{i})\,,
\label{eq:umdpd}
\end{equation}
where the one-body terms depend on local densities,
\begin{equation}
\rhobar_i={\textstyle\sum_{i\ne j}}\,\wrho(r_{ij})\,.
\label{eq:rhoi}
\end{equation}
From the potential one can derive the force law,
\begin{equation}
\Fvec_i=-\frac{\partial U}{\partial\rvec_i}
={\textstyle\sum_{i\ne j}}\,\Fvec_{ij}\,,
\label{eq:fij}
\end{equation}
where
\begin{equation}
\Fvec_{ij}=-[u'(\rhobar_i)+u'(\rhobar_j)]\,
\wrho'(r_{ij})\,\hatrvec_{ij}\,.
\label{eq:mdpd}
\end{equation}
Here $\wrho(r)\ge0$ is another weight function, also with compact
support.  Since the potential in Eq.~\eqref{eq:umdpd} is a regular
function of the particle positions, MDPD avoids issues that otherwise
plague density-dependent pair interactions \cite{Lou02}, although if
the forces are not purely repulsive one should take care to ensure
thermodynamic stability according to the criteria devised by Ruelle
\cite{Rue99}.  

It is clear that the choice $u(\rhobar)=A\rhobar/2$ and
$\wrho'(r)=-w_C(r)$ brings Eq.~\eqref{eq:mdpd} into agreement with
Eq.~\eqref{eq:dpd}.  Hence standard DPD is just a special case of
MDPD.  Note that this may imply $\int\! d^3\rvec\,\wrho(r)\ne1$ but
abandoning this normalisation requirement leads to a considerable
notational simplification by eliminating unnecessary prefactors.
Another example is $u(\rhobar)=B\rhobar^2/2$.  Again setting
$w_C=-\wrho'$, this generates the force law
\begin{equation}
\Fvec_{ij}=B\,(\rhobar_i+\rhobar_j)\,w_C(r_{ij})\,\hatrvec_{ij}\,.
\label{eq:b}
\end{equation}
This force law (with $B>0$) in combination with the standard DPD force
law of Eq.~\eqref{eq:dpd} (with $A<0$ and a larger cut-off) has
been extensively used for free surface simulations.  For a recent
review see Ghoufi \etal\ \cite{GEM13}

Frequently DPD is applied to multicomponent systems and
Eq.~\eqref{eq:dpd} is generalised to
\begin{equation}
\Fvec_{ij}=A_{ij}\,w_C(r_{ij})\,\hatrvec_{ij}
\label{eq:mcdpd}
\end{equation}
where $A_{ij}$ is a matrix of repulsion amplitudes.  It is natural to
consider a similar generalisation of Eq.~\eqref{eq:b},
\begin{equation}
\Fvec_{ij}=B_{ij}\,(\rhobar_i+\rhobar_j)\,w_C(r_{ij})\,\hatrvec_{ij}\,.
\label{eq:bij}
\end{equation}
This has been proposed in the published literature \cite{APL+11,
  SAM+13}, but my claim is that such a force law is \emph{not}
conservative unless $B_{ij}$ is a constant matrix.  This is the no-go
theorem of the title. (In fairness to the authors of
Refs.~\cite{APL+11, SAM+13}, they actually only use the $B_{ij}=B$
case.)

How can the no-go theorem be proved?  Hopefully it is obvious that a
\emph{sufficient} condition for a many body force law to be
conservative is to display an explicit potential.  For the present
problem, if $B_{ij} = B$, such a potential is provided by
Eqs.~\eqref{eq:umdpd} and \eqref{eq:rhoi} with $u(\rhobar) =
B\rhobar^2/2$ as already stated.  A \emph{necessary} condition for a
force law to be conservative is that the `Maxwell relation'
\begin{equation}
\frac{\partial\Fvec_i}{\partial\rvec_j}=
\frac{\partial\Fvec_j}{\partial\rvec_i}
\label{eq:maxwell}
\end{equation}
is satisfied.  This follows from the first part of Eq.~\eqref{eq:fij}.
The application of this to a general configuration of $N$ particles is
not straightforward since both sides of Eq.~\eqref{eq:maxwell} contain
multiple sums.  However Eq.~\eqref{eq:maxwell} should hold for
\emph{any} configuration of particles, so we can choose a
configuration at our convenience.  For the present problem therefore,
let us select one which contains an isolated collinear triplet of
particles.  Without loss of generality we can place the particles on
the $x$-axis at positions $x_1<x_2<x_3$.  I shall define
$x_{ij}=x_j-x_i$.  We can further suppose $x_{13}<r_c$ so that all
particles interact.  The pairwise forces are (setting $w_C=-\wrho'$)
\begin{equation}
\begin{array}{l}
F_{12}=-F_{21}=-B_{12}\,(\rhobar_1+\rhobar_2)\,\wrho'(x_{12})\,,\\[3pt]
F_{13}=-F_{31}=-B_{13}\,(\rhobar_1+\rhobar_3)\,\wrho'(x_{13})\,,\\[3pt]
F_{23}=-F_{32}=-B_{23}\,(\rhobar_2+\rhobar_3)\,\wrho'(x_{23})\,.
\end{array}
\end{equation}
The local densities are $\rhobar_1 = \wrho(x_{12}) + \wrho(x_{13})$,
$\rhobar_2 = \wrho(x_{12}) + \wrho(x_{23})$, $\rhobar_3 =
\wrho(x_{13}) + \wrho(x_{23})$; and the total forces are $F_1 = F_{12}
+ F_{13}$, $F_2 = F_{21} + F_{23}$, $F_3 = F_{31} + F_{32}$.  Let us
define $D_{ij} = {\partial F_i}/{\partial x_j} - {\partial
  F_j}/{\partial x_i}$.  By explicit calculation I find
$D_{12}=D_{23}=D_{31}$ where
\begin{equation}
\begin{array}{l}
D_{12}=(B_{13}-B_{23})\,\wrho'(x_{13})\,\wrho'(x_{23})\\[6pt]
\hspace{4em}{}+(B_{12}-B_{13})\,\wrho'(x_{12})\,\wrho'(x_{13})\\[6pt]
\hspace{6em}{}+(B_{12}-B_{23})\,\wrho'(x_{12})\,\wrho'(x_{23})\,.
\end{array}
\end{equation}
Thus we see the $D_{ij}$ vanish if and only if $B_{12}=B_{13}=B_{23}$,
since by choice all the $\wrho'(x_{ij})$ factors are strictly
negative.  Moreover we can pick \emph{any} three particles for this
argument.  Hence the Maxwell relation fails in at least in a
\emph{subset} of configurations, unless $B_{ij}$ is a constant matrix.
This completes the desired proof of necessity.

An analogous argument does not go through for Eq.~\eqref{eq:mcdpd}
since in that case we can exhibit an actual potential, namely
$U=\sum_{i>j}A_{ij}\wrho(r_{ij})$, although this does not reduce to a
sum of self energies unless $A_{ij}=A_i+A_j$.  Of course one can
additively combine force laws, so it is possible to have conservative
multicomponent MDPD based on Eqs.~\eqref{eq:b} and \eqref{eq:mcdpd}.
Other approaches have been described by
Trofimov \etal\ \cite{TNM02} and Merabia \etal\ \cite{MBP08}.

Why is it so important that the force law be conservative?  The answer
is that for many applications one wishes to use the machinery of
equilibrium statistical mechanics and thermodynamics \cite{WVH+98},
which requires the existence of a potential so that the steady state
is given by Gibbs-Boltzmann.  Of course the absence of a potential
does not preclude the existence of a \emph{non}-equilibrium steady state.
Hence the effects of a non-conservative force law could be quite
subtle, a bit like a failure to satisfy detailed balance in a
Monte-Carlo simulation \cite{FS02}.

Here the general strategy to prove the no-go theorem has been to
identify a convenient particle configuration for which the sums
implicit in Eq.~\eqref{eq:maxwell} become manageable.  Presumably this
could be applied in other cases too, but it would seem that at least
$N\ge3$ particles are required since a central force between a pair of
particles can always be integrated to a pair potential.  The converse
implication is that a similarly defined potential for assembling
$N\ge3$ particles would depend on the assembly path, unless
Eq.~\eqref{eq:maxwell} holds.


%

\end{document}